# Study of the optical properties and formation process of nanocrystalline graphite layers on silicon substrates

Z.E. Kun'kova[1]*, O.V. Vikhrova[2], Yu.A. Danilov[2], A.I. Rukovishnikov[1], A.A. Fedorova[3], Yu.V. Markin[1], R.N. Kryukov[2], V.P. Lesnikov[2], A.V. Nezhdanov[2], P.A. Yunin[4]

[1] V.A. Kotelnikov Institute of Radio Engineering and Electronics, Russian Academy of Sciences, Fryazino branch, Fryazino, Russia

[2] Lobachevsky State University, Nizhny Novgorod, Russia

[3] V.A. Kotelnikov Institute of Radio Engineering and Electronics, Russian Academy of Sciences, Moscow, Russia

[4] Institute for Physics of Microstructures, Russian Academy of Sciences, Nizhny Novgorod, Russia

*z.kunkova@gmail.com, danilov@nifti.unn.ru

## Abstract

The influence of temperature (200–500°C) and deposition time (2–15 min) on the optical properties and surface topography of carbon layers formed by pulsed laser deposition on Si substrates was investigated. The techniques employed included spectral ellipsometry, Raman spectroscopy, atomic force microscopy, and X-ray reflectometry. It was found that the thickness and spectra of the ellipsometry parameters of the layers deposited for 2 min at substrate temperatures of 200, 300, and 400°C are very similar; however, the surface topography of the layers depends on the substrate crystallographic orientation. At 500°C, the ellipsometry spectra and surface roughness characteristics of the layers change with the deposition time, owing to an increase in the thickness and the specific nature of the layer-formation processes. It was found that the layers produced at 500°C consist of nanocrystalline graphite—specifically, 4–5-layer graphene with a grain size of ~7 nm. For the layers deposited at 500°C for 10 and 15 min, the spectra of the optical constants $n(\lambda)$ and $k(\lambda)$ were calculated for the wavelength range $\lambda$ = 300–1000 nm.

## 1. Introduction

Graphene has attracted increasing interest of researchers due to its unique electronic, thermal, mechanical, and chemical properties since its production and initial researches. Applications of graphene in field-effect transistors, supercapacitors, sensors, protective coatings, flexible transparent displays, solar cells, and other applications have been proposed [1-3], but its production and integration into devices have proven complex and expensive. Thin layers of other $sp^2$-carbon morphologies are being actively studied as an alternative to graphene layers. These thin layers possess a number of attractive physical and chemical properties and offer lower production costs. Numerous methods for forming such materials have been developed and are being refined, aimed at gaining fundamental knowledge and industrial implementation. Thin layers of nanocrystalline graphite/graphene (NCG), which are an aggregate of $sp^2$-linked nanodomains of varying shapes and orientations, are among the materials being actively studied. NCG layers have such useful properties as good electrical and thermal conductivity, optical transparency, mechanical and chemical strength and they are already used in field-effect transistors, sensors, supercapacitors and protective coatings [3, 4].

NCG layers are obtained by various methods: chemical synthesis [5], pyrolysis [6], atomic layer deposition (ALD) [7], molecular beam epitaxy (MBE) [8], chemical vapor deposition (CVD) [9-11], laser deposition (PLD) [4, 12-17], etc. In the most cases, the NCG deposition requires complex technological equipment and hard process conditions, including high temperatures (~ 1000°C), high vacuum, and toxic precursors. Currently, CVD, in particular PECVD (plasma-enhanced chemical vapor deposition), which allow for a reduction in the substrate temperature, thereby expanding the range of possible applications, are the most studied and used methods.

Pulsed laser deposition (PLD) is considered as an alternative to CVD methods. The advantages of PLD methods include their relative simplicity and flexibility: lower substrate temperature and lower vacuum; the ability to introduce dopants into the growing film; the ability to use background gases and implement chemical interactions between the removed particles and the background gas, etc. There are two main approaches to PLD of graphene and graphene-like films: 1 - using metal catalysts deposited on a substrate; 2 - without catalysts. The first approach is more common, as it allows for the production of graphene-like films with fewer layers and defects. The second approach is attractive because it allows for the growth of layers directly on desired substrates. However, the latter has not been sufficiently studied, and further research is needed to find optimal PLD conditions without the use of metal catalysts.

In this study, we investigated the effects of the deposition temperature and time on the optical properties and surface topography of thin carbon layers on Si substrates formed by low-temperature PLD (200–500°C) without catalysts. We are not aware of any publications characterizing graphene-like layers deposited by PLD using spectral ellipsometry and atomic force microscopy.

## 2. Samples and Research Methods

We studied carbon layers formed by the PLD method in vacuum at pressure of $< 10^{-6}$ Torr. A pyrolytic graphite plate was the target, and an LQ 529 YAG:Nd laser operating at the second harmonic (wavelength 532 nm), with a pulse duration of 10 ns, a repetition rate of 10 $s^{-1}$, and a pulse energy of 250 mJ was the radiation source. The laser beam was focused onto the target to form a spot with an area of ~1.5 $mm^2$. Single-crystal silicon wafers, Si(100), Si(111) and other treated by boiling in toluene and acetone, followed by rinsing in deionized water, were the substrates.

The optical properties of the carbon layers were studied using spectral ellipsometry (SE) and Raman spectroscopy. The spectra of ellipsometric angles, $\Psi(\lambda)$ and $\Delta(\lambda)$[1], were recorded in the wavelength range $\lambda$ = 300–2200 nm with a resolution of ~ 7 nm. The light spot size was ~ 0.4 ×1 mm. The LED spectroellipsometer with binary modulation of the polarization state, developed at the V.A. Kotelnikov Institute of Radio Engineering and Electronics, RAS (Fryazino Branch) [18], was used. The Raman spectra were studied with an NTEGRA Spectra complex (NT-MDT) using a laser with a wavelength of 473 nm. The radiation was focused with the use of a 100× objective with aperture of NA = 0.9. The laser power was 0.5 mW. Raman spectra were studied in the backscattering geometry in the range of 50–3200 $cm^{-1}$ with a resolution of 0.7 $cm^{-1}$. All spectra were recorded at room temperature.

The surface topography of the layers was studied using an atomic force microscope SmartSPM (AIST-NT). The thickness, roughness, and density of carbon layers in several structures were determined using X-ray diffraction reflectometry (XRD) (a Brucker D8 Discover diffractometer).

## 3. Results and discussion

In this work, a series of S1–S7 structures formed by varying the Si substrate temperature, $T_g$, in the range of 200–500°C and the deposition time of carbon layer, $t_g$, from 2 to 15 min were investigated. Table 1 shows the designations of the C-layers in the S1–S7 structures, their deposition regimes, and XRD, spectral ellipsometry (SE), and AFM characterization data.

Table 1. $T_g$ is the substrate temperature, $t_g$ is the deposition time; XRD data – $d$ is the thickness of a C-layer, $R$ is the roughness height, $D$ is the layer density; SE data – $d_{el}$ are the

---

[1] Ellipsometric angles $\Psi(\lambda)$ and $\Delta(\lambda)$ characterize the relative changes in amplitude and phase experienced during reflection of the $p$- and $s$-components of the electric vector of a light wave [19].

thicknesses of C- and $SiO_2$-layers; AFM data – *S*q is the root-mean-square roughness, *S*z is the maximum roughness height.

| 1 | 2 | | 3 | 4 | | 5 | |
|---|---|---|---|---|---|---|---|
| C-layer N | $T_g$ | $t_g$ | XRD | $d_{el}$, nm | | $S_q$ | $S_z$, |
| | | | | C | $SiO_2$ | nm | |
| C-S1 | 200°C | 2 min | $d$ = 13.3 nm, $R$ = 1.7 nm, $D$ = 2.85 g·cm$^{-3}$ | 11 | 6 | 1.172 | 39.08 |
| C-S2 | 300°C | 2 min | $d$ = 13.7 nm, $R$ = 1.7 nm, $D$ = 2.86 g·cm$^{-3}$ | 11 | 6 | 1.039 | 57.62 |
| C-S3 | 400°C | 2 min | – | 13 | 9 | 1.443 | 11.04 |
| C-S4 | 500°C | 2 min | $d$ = 22.9 nm, $R$ = 5.9 nm, $D$ = 3.09 g·cm$^{-3}$ | 33 | 7 | 0.226 | 1.93 |
| C-S5 | 500°C | 5 min | – | 45 | 8.4 | 1.093 | 59.86 |
| C-S6 | 500°C | 10 min | – | 117 | 3 | 0.373 | 12.4 |
| C-S7 | 500°C | 15 min | – | 180 | 7 | 0.533 | 12.77 |

The spectra of the refractive index and extinction coefficient, $n(\lambda)$ and $k(\lambda)$, as well as the thicknesses of the C- and $SiO_2$-layers, were calculated using a three-layer model (semi-infinite Si-substrate/$SiO_2$layer/C-layer) by fitting to the measured $\Psi(\lambda)$ и $\Delta(\lambda)$ spectra via the least-squares method, employing standard ellipsometric equations and tabulated optical constants for Si and $SiO_2$ [20].

The spectra of the ellipsometric parameters $\Psi(\lambda)$ and $\Delta(\lambda)$ were recorded for the angles of light incidence $\varphi$ = 60, 70, and 75°. Fig. 1(a, b) shows by symbols the spectra of structures S2–S7 for $\varphi$ = 70°. In the spectral range studied, the spectra $\Psi(\lambda)$ and $\Delta(\lambda)$ of structures S1, S2, and S3 are very close, so the spectra for S1 structure are not shown. Fig. 1(a, b) also shows the spectra $\Psi(\lambda)$ and $\Delta(\lambda)$ for bulk silicon, calculated using the literature data [20]. The results of modeling the $\Psi(\lambda)$ and $\Delta(\lambda)$ dependences are shown in Fig. 1(a, b) by solid lines, and good agreement between the experimental and calculated spectra (maintained for the other two incidence angles) can be seen.

In Fig. 1(a, b), the $\Psi(\lambda)$ and $\Delta(\lambda)$ spectra for all the structures exhibit peaks and/or inflections in the range $\lambda$ < 400 nm, which correspond to characteristic features in the spectra of silicon and arise from the reflection at the substrate front surface. Consequently, the thicknesses of all the C-layers under investigation are insufficient to absorb the incident radiation. As a result, the detected signal in addition to the radiation reflected by the C-layer and the Si-substrate front surface may include a contribution from the substrate back surface, which is not accounted for in the three-layer model.

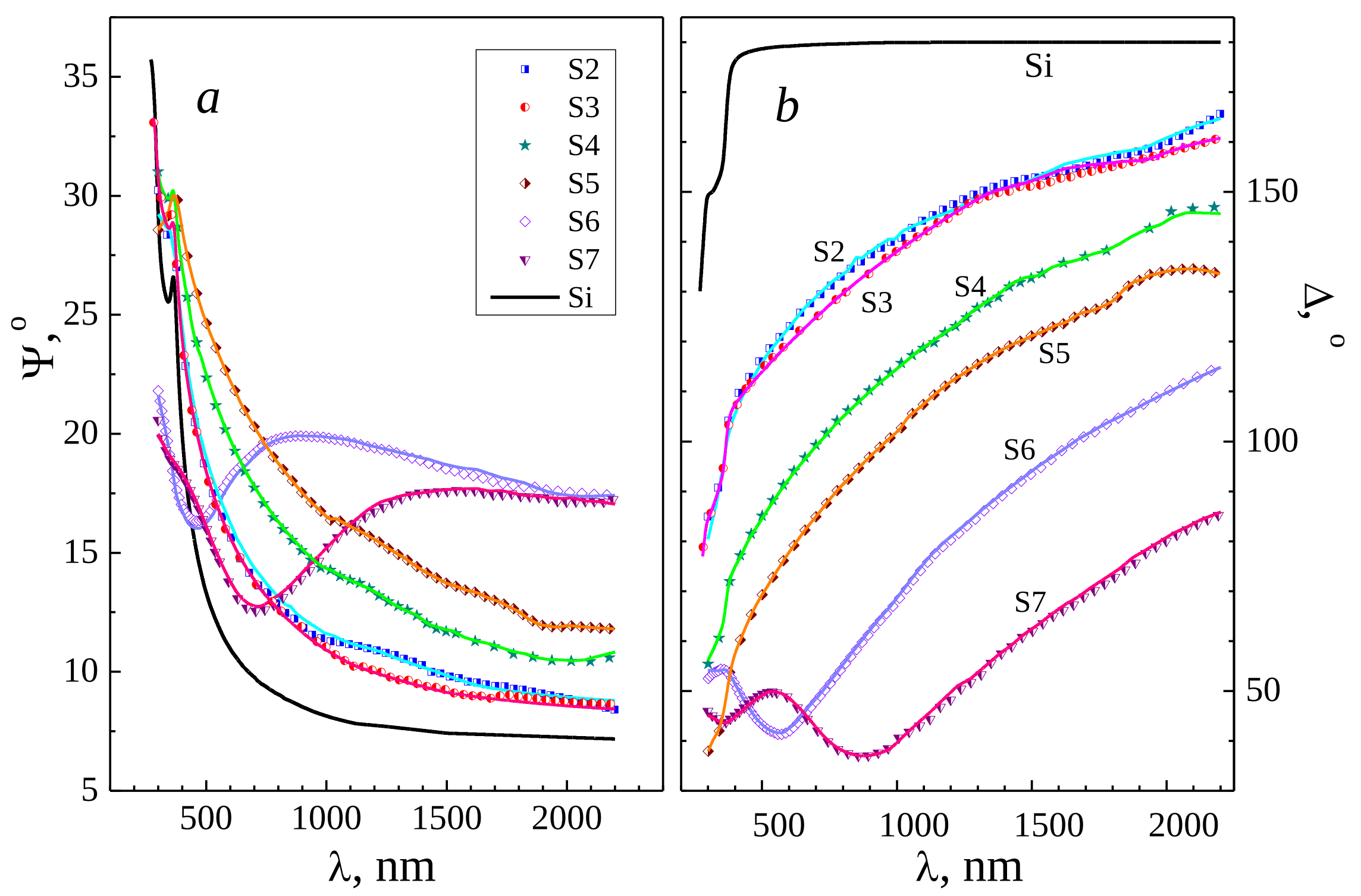

Fig. 1. Measured (symbols) and calculated (solid lines) spectral dependences of the ellipsometric angles Ψ(λ) (*a*) and Δ(λ) (*b*) for the S2–S7 structures, φ = 70°.

A comparison of the calculated *n*(λ) and *k*(λ) dependencies for the C-layers in the S1–S7 structures with the published spectra of the optical constants for the NCG layers formed by other methods [21] revealed significant discrepancies; this can naturally be attributed to an unwanted signal reflected from the Si-substrate back surface. Результаты сравнения позволяют считать корректными спектры *n*(λ) и *k*(λ) слоёв C-S6 и C-S7 в области λ = 300–1000 nm, а также оценки толщины $d_{el}$, приведённые в колонке 4 Таблицы 1. For the C-S1–C-S5 layers, the differences in the spectra are significant across the entire measurement range, while for the C-S6 and C-S7 layers they are substantial in the region where silicon is transparent (λ > 1000 nm). The comparison results show that the *n*(λ) and *k*(λ) spectra of the C-S6 and C-S7 layers in the range λ = 300–1000 nm, as well as the thickness estimates, $d_{el}$, for all layers (column 4 in Table 1) are correct.

The calculated spectral dependencies *n*(λ) and *k*(λ) for the C-S6 and C-S7 layers are presented in Fig. 2(a, b). We compared them with the data published in [21], where the optical properties of nanocrystalline graphite (NCG) layers deposited via PECVD on $SiO_2$/Si substrates ($T_g$ = 900°C, $t_g$ = 1–60 min) were investigated in detail. In the region of the intrinsic substrate absorption (λ < 1000 nm), the $n$(λ) and $k$(λ) curves for the C-S6 layer repeat the shape of the comparison spectra in [21]. The $n$ and $k$ values in this region are close to the optical constants of the NCG layer formed at $T_g$ = 900°C, $t_g$ = 30 min ($d$ ~ 83 nm).

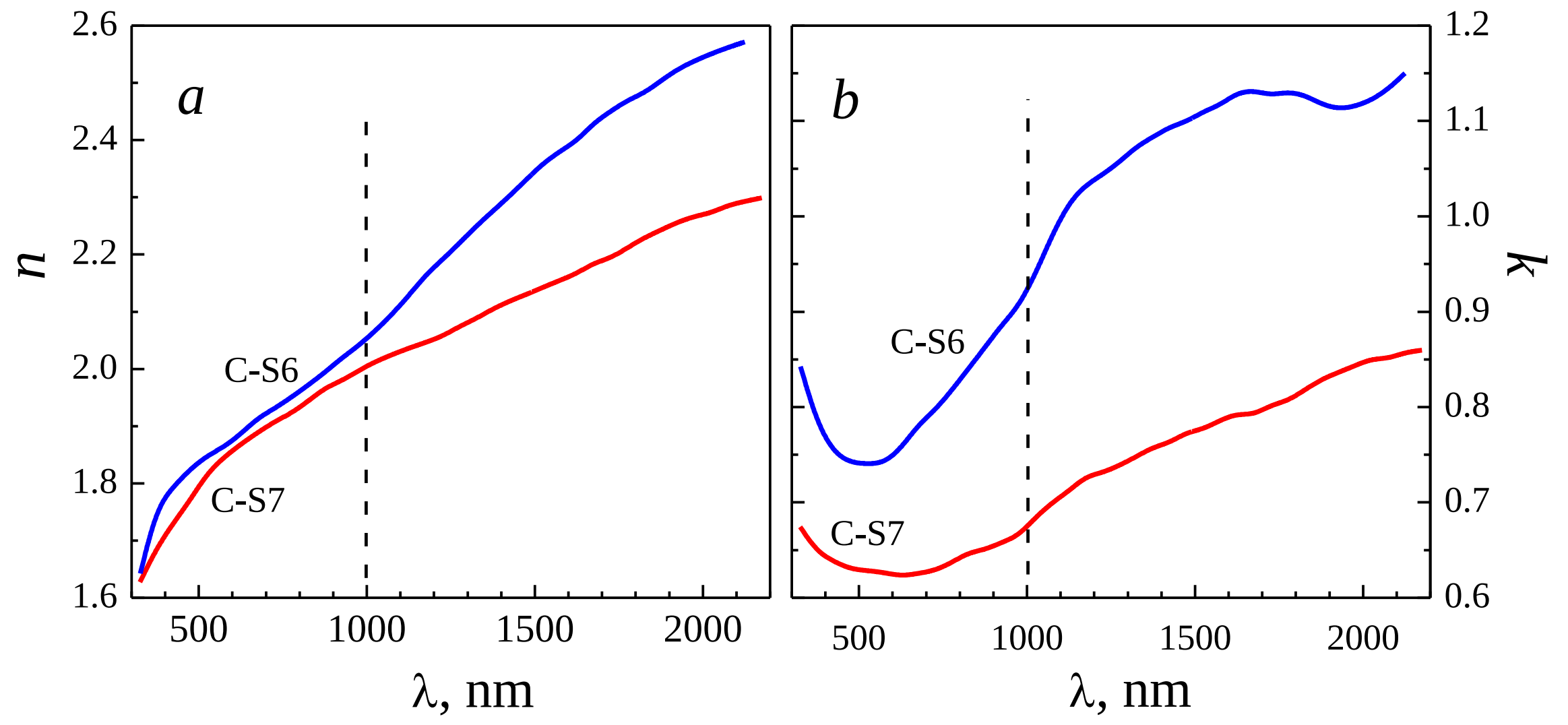

Fig. 2. Calculated spectra $n$(λ) ($a$) and $k$(λ) ($b$) for the C-layers in the S6 and S7structures.

In the transparency range of the Si substrate (λ > 1000 nm), a sharper increase in the $n$ and $k$ values than in the spectra [21] is observed, which is naturally to associate with an unwanted signal. In the spectra of the C-S7 layer ($d_{el}$ ~ 180 nm), the contribution of this signal is insignificant, but the characteristic spectral features are broadened and less pronounced. These features broadening may be caused by an increase in the content of amorphous carbon and defects in the C-S7 layer.

Raman spectroscopy was used to elucidate the structure of the studied C-layers. Two ranges can be distinguished in the experimental Raman spectra: the semiconductor region (up to 1100 cm$^{-1}$) and the carbon region (from 1100 to 3200 cm$^{-1}$). The semiconductor region contains an intense narrow peak at ~521 cm$^{-1}$, corresponding to single-crystal Si, and a less intense satellite in the 950–1000 cm$^{-1}$ region, which are due to insufficient carbon film thickness.

The carbon part of the spectrum of the studied structures is shown in Fig. 3. Here, a system of characteristic peaks of carbon structures is observed: at ~ 1370 (peak D), ~ 1600 (peak G), 2740 (peak 2D), and 2950 cm$^{-1}$ (peak D+D'). Peak G is associated with vibrations of carbon atoms with $sp^2$-bonds in the graphene plane. Peak D corresponds to defects in the graphene lattice. The presence of peaks 2D (an overtone of peak D) and D+D' is also characteristic of graphene and graphene-like structures. [22, 23].

Fig. 3 shows that with increasing thickness of the carbon layers (spectra C-S4, C-S5, C-S7) deposited at $T_g$ = 500°C, the intensity of the Raman peaks increases, peaks G and D narrow, and at the thickness maximum studied, peaks 2D and D+D' are well resolved. The Raman spectra obtained for layers deposited at 500°C are characteristic of nanocrystalline graphite (NCG).

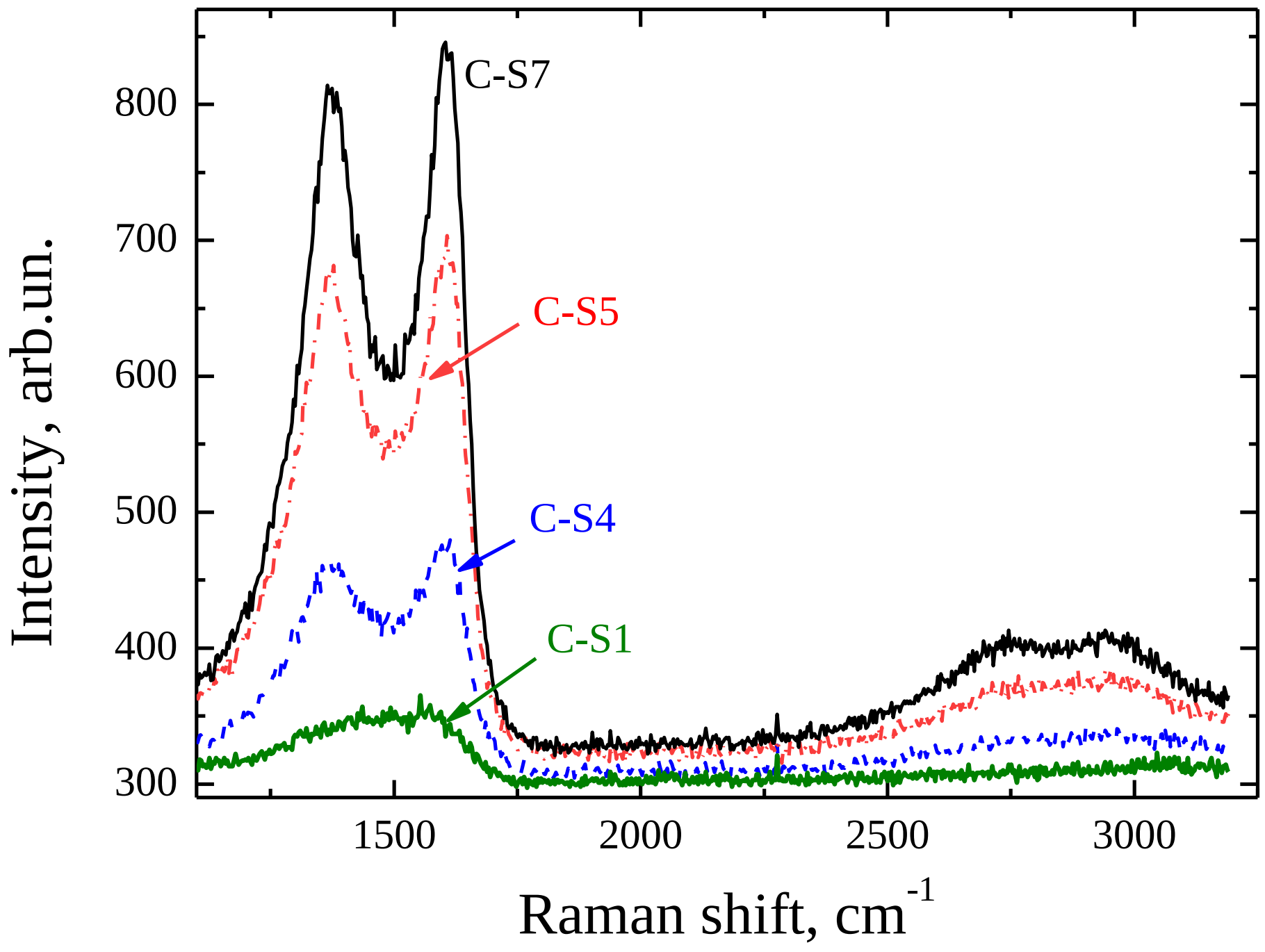

Fig. 3. Raman spectra for structures with C-layers deposited at $T_g$ = 500 (curves C-S4, C-S5, C-S7) and 200°C (curve C-S1).

As the deposition temperature decreases only the spectrum of the C-S1 layer (Fig. 3) shows, as the spectra of the C-S2 and C-S3 layers are virtually indistinguishable from it), the intensity of the G and D peaks decreases, they broaden, and merge, possibly indicating the formation of amorphous layers. The 2D and D+D' peaks completely disappear, indicating a loss of long-range order.

To obtain some quantitative characteristics of the C-layers, a procedure for decomposing the carbon spectral peak system into Lorentzians was used. Fig. 4 shows the fit of the Raman spectrum peaks for the C-S7 layer. Good agreement between the experimental and fitted spectra is evident. Table 2 presents the obtained Raman peak characteristics.

Table 2. Fitting parameters for the Raman peaks of the C-S7 layer: $x_c$ - peak position; $w$ - peak full width at the half-maximum; $A$ - area under the peaks; $H$ - peak height.

| | peak D | peak G – | peak 2D | peak D+D' |
|---|---|---|---|---|
| $x_c$, cm$^{-1}$ | 1378.7±1.1 | 1599.4±0.7 | 2723±14 | 2966±15 |
| $w$, cm$^{-1}$ | 165.9±3.6 | 89.2±2.4 | 220±43 | 266±48 |
| $A$, arb. un. | 118600±2000 | 67600±1500 | 19800±4800 | 26900±5400 |
| $H$, arb. un. | 455.0±0.7 | 482±8 | 57±7 | 64±5 |

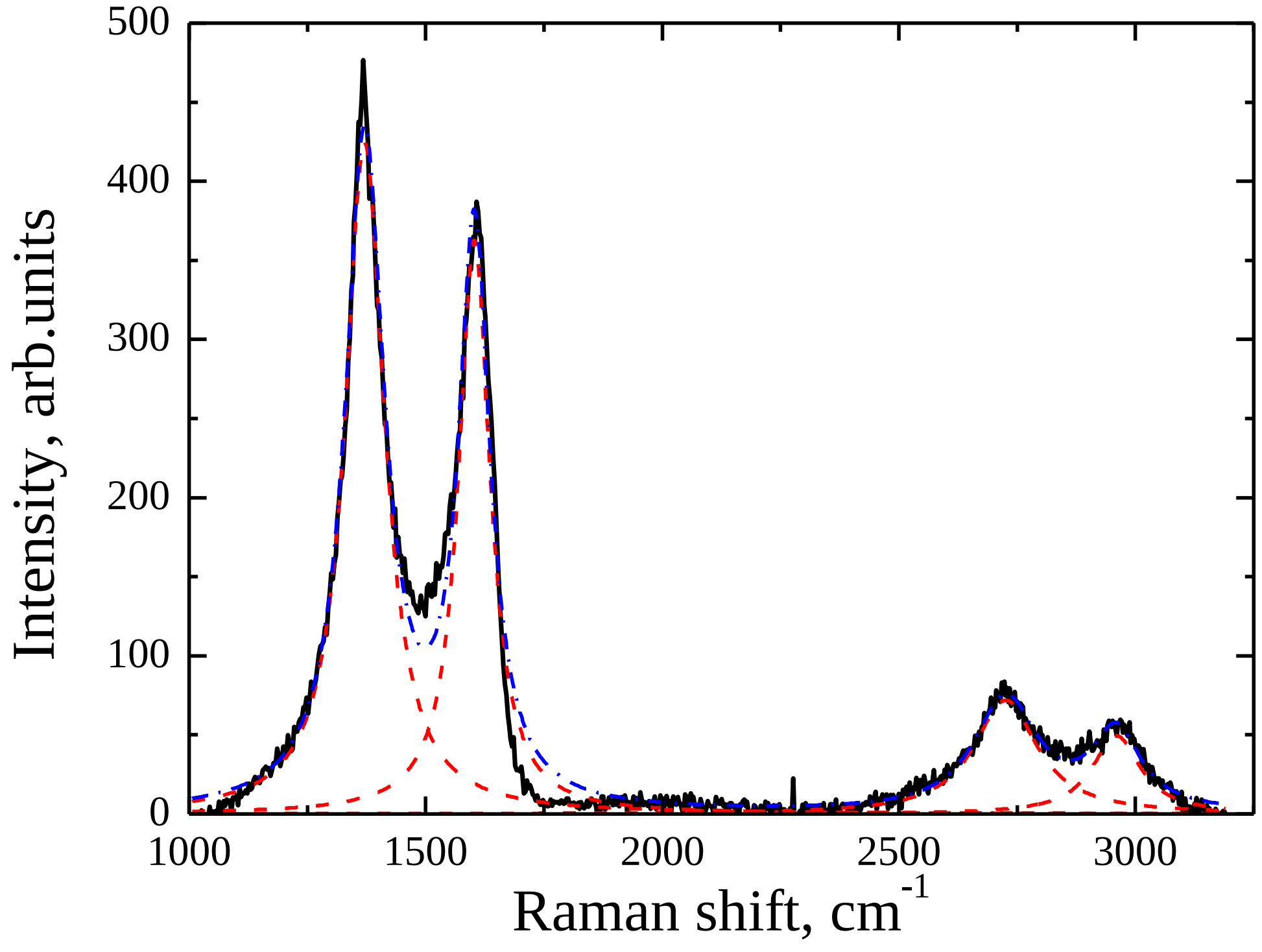

Fig. 4. Raman spectrum of the C-S7 layer. The experimental spectrum is the solid black line, the fit for the four peaks with Lorentzians under background-subtracting is the dashed red lines, and the sum of the Lorentzians is the dashed-dotted blue line.

In [23], it was shown that the G and 2D Raman peaks change their shape, position, and relative intensity depending on the number of graphene layers. In [23], the full width at the half maximum ($x_c$ in Table 2) of the 2D peak is given: for three-layer graphene, $x_c$ = 74 cm$^{-1}$, for few-layer graphene, 95 cm$^{-1}$, and for 7–8-layer graphene, 402 cm$^{-1}$. For the C-S7 layer, the $x_c$ value is 220 cm$^{-1}$, which corresponds to five-layer graphene. For thinner C-layers in the S4 and S5 structures, this value remains virtually unchanged.

The planar grain size (coherence region) can be calculated using the formula [22]

$$L_a = 2.4 \cdot 10^{-10} \cdot \lambda^4 A_G / A_D ,$$

where λ [nm] is the wavelength of the excitation laser (in this work, λ = 473 nm), $A$ is the area under the corresponding peak, and $L_a$ is the grain size in nanometers. For the C-S7 layer, the calculated value of $L_a$ is ~ 7 nm. The grain size estimates for the C-layers in the structures S4 and S5 are essentially the same. The differences in the Raman spectra of the C-layers in the structures S4, S5, and S7 are apparently due to only a decrease in the signal intensity with decreasing C-layer thickness.

The formation of C-layers under different growth conditions was studied using atomic force microscopy (AFM). The lower row in Fig. 5(a–d) shows AFM scans of the surface fragments of the C-S1–C-S4 layers obtained at the deposition time $t_g$ = 2 min and substrate temperatures $T_g$ of 200, 300, 400, and 500°C. The scan area size is approximately 4 × 4 µm. The top row shows 3d-images of the same surface areas of the layers, with the exception of the C-S2 layer, for which a more informative 3d-image of approximately 10 × 10 µm is given.

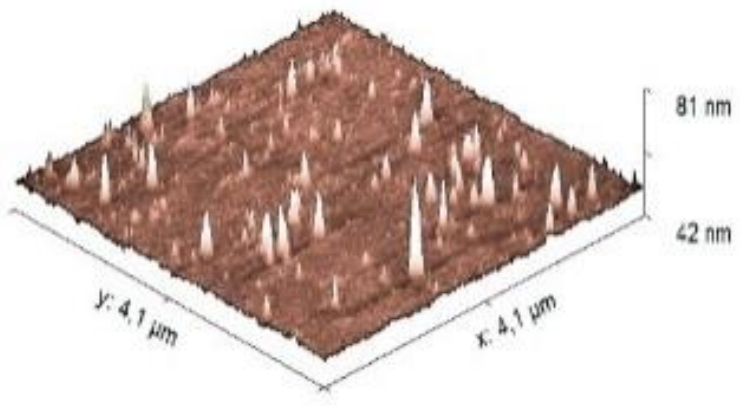

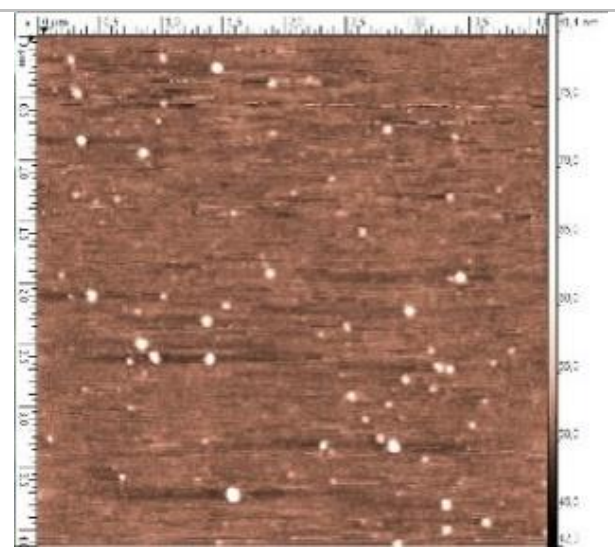

(a) C-S1 на Si(100) (200°C, 2 min)
~ 4.1 × 4.1 µm

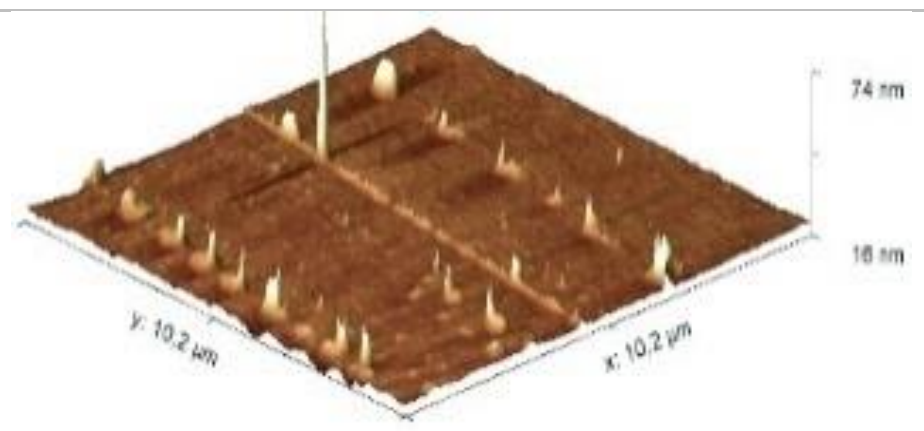

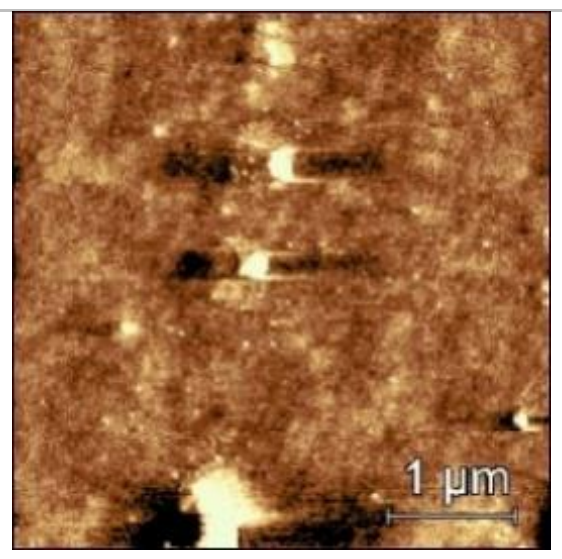

(b) C-S2 на Si(111) (300°C, 2 min)
~ 4.0 × 4.0 µm
(3d ~ 10 × 10 µm)

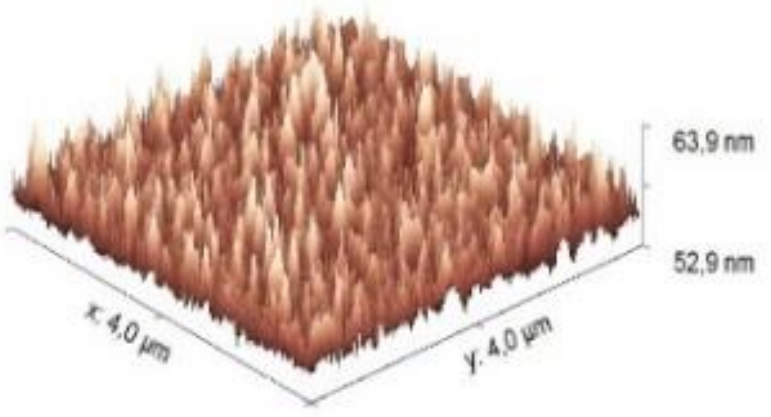

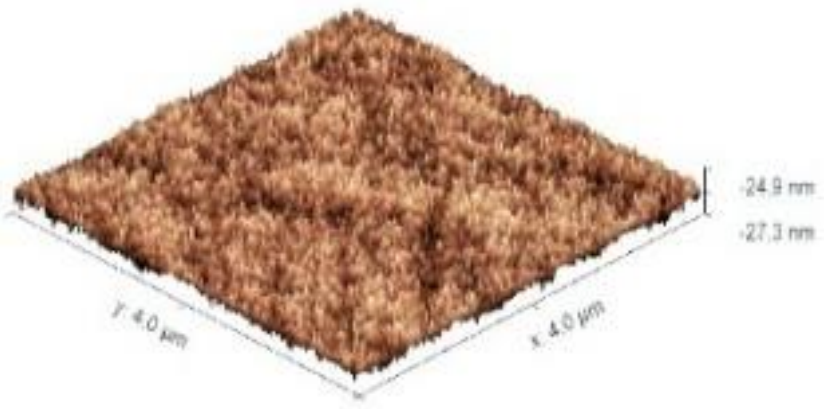

| | |
|---|---|
| 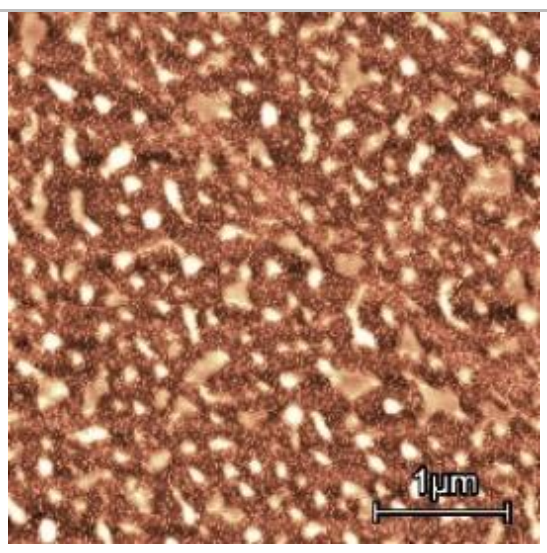 | 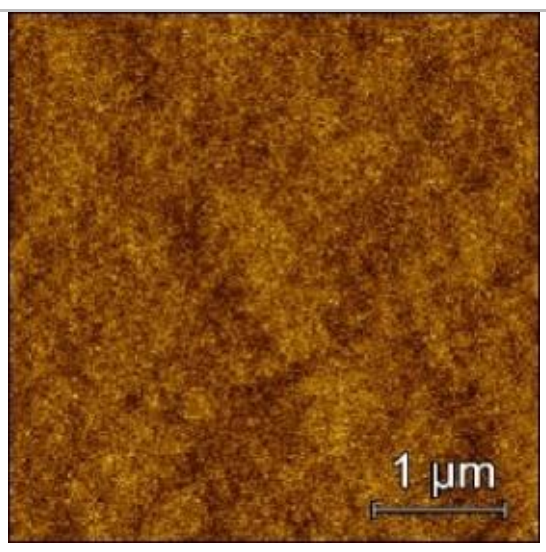 |
| (c) C-S3 на Si(100) (400ºC, 2 min)<br>~ 4.0 × 4.0 µm | (d) C-S4 на Si(111) (500ºC, 2 min)<br>~ 4.0 × 4.0 µm |

Fig. 5. (a, b) Bottom row: AFM scans of the C-S1 and C-S2 layers surfaces with a size of ~ 4 × 4 µm. (a) Top row: 3d-image of the same surface area of the C-S1 layer; (b) 3d-image of the C-S2 layer surface with a size of ~ 10 × 10 µm.
(c, d): Bottom row - 2d and top row - 3d-AFM images of surface fragments of the layers C-S3 and C-S4 with a size of ~ 4 × 4 µm.

The 2d and 3d images of the C-S1 layer, Fig. 5(a), show relatively rare "peaks" with the maximum height $h_{max}$ ~ 40 nm. They can be associated with the deposition of large droplets of the evaporated material. Possible small nucleation islands are not resolved due to the presence of the high "peaks" on the surface. The layer thickness estimates for C-S1, $d_{ell}$ ~ 11 nm and $d_{XRD}$ ~ 13.3 nm, appear to correspond to the average roughness height. In the 3d-image of the C-S2 layer (area of ~10 × 10 µm, Fig. 5(b)), parallel stripes space apart d ~ 3–5 µm are clearly visible. These can be associated with steps on the vicinal Si(111) substrate. The density of the 3D islands-nuclei is low with the maximum height $h_{max}$ ~ 10 - 20 nm (an exception is the single "peak" with $h_{max}$ ~ 59 nm). In the AFM images of the C-S3 layer surface, Fig. 5(c), numerous small nuclei are clearly visible, while no high "peaks" are observed. The $h_{max}$ value for these nuclei is approximately 11 nm, which is close to the C-layer thickness estimate $d_{ell}$ ~ 13 nm. The 2d-images show the coalescence of these nuclei and an increase in their lateral dimensions. It appears that under the ablation regime employed the deposition conditions $T_g$ = 400°C and $t_g$ = 2 min facilitate the formation of the nuclei. In the 2d- and 3d-images of the layer S-S4, Fig. 5 (d), there are no formations in the form of "peaks" and/or small initial islands-nuclei: the layer C-S4 surface is fine-grained with a slight roughness. It can be assumed that under the conditions $T_g$ = 500°C, $t_g$ = 2 min, an amorphous C-layer with a thickness ~ 20–30 nm and $h_{max}$ ~ 2 nm is formed.

AFM images of the surface fragments (~ 4 × 4 µm in size) of the C-S5–C-S7 layers deposited at the same growth temperature, $T_g$ = 500 °C, and deposition times $t_g$: a – 5; b – 10; c - 15 min, are presented in Fig. 6 (a - c). A more detailed surface relief for the scan areas

~ 1 × 1 μm layers C-S6 and C-S7 is shown in Fig. 7 (a, b).

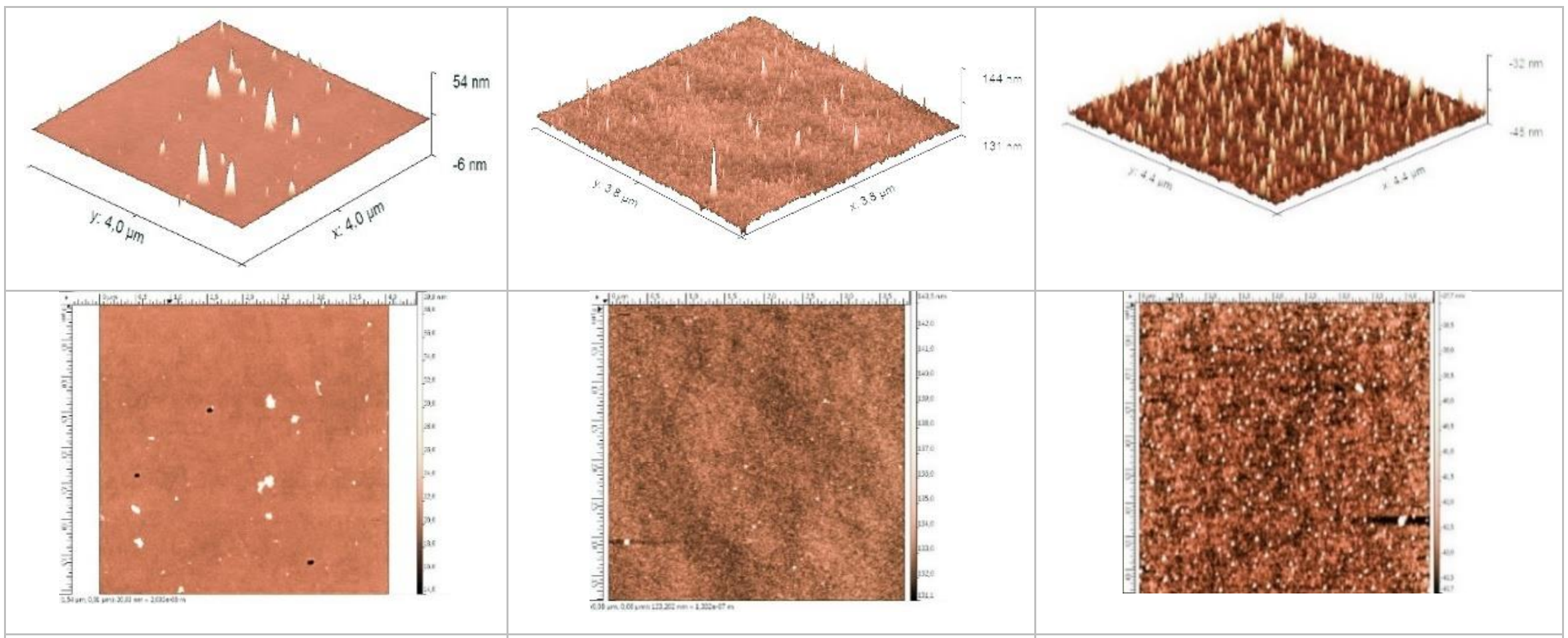

| | | |
|---|---|---|
| (a) C-S5 на Si(1110-KES) (500ºC, 5 min) ~ 4.0 × 4.0, μm | (b) C-S6 на Si(100) (500ºC, 10 min) ~ 3.8 × 3.8, μm | (c) C-S7 на Si(1110-KES) (500ºC, 15 min) ~ 4.4 × 4.4, μm |

Fig. 6(a–c). Bottom row: AFM images of surface fragments of the layers C-S5–C-S7, with dimensions of ~ 4 × 4 μm. $T_g$ = 500°C; $t_g$: (a) 5, (b) 10, (c) 15 min. Top row: 3d-images of the same surface areas.

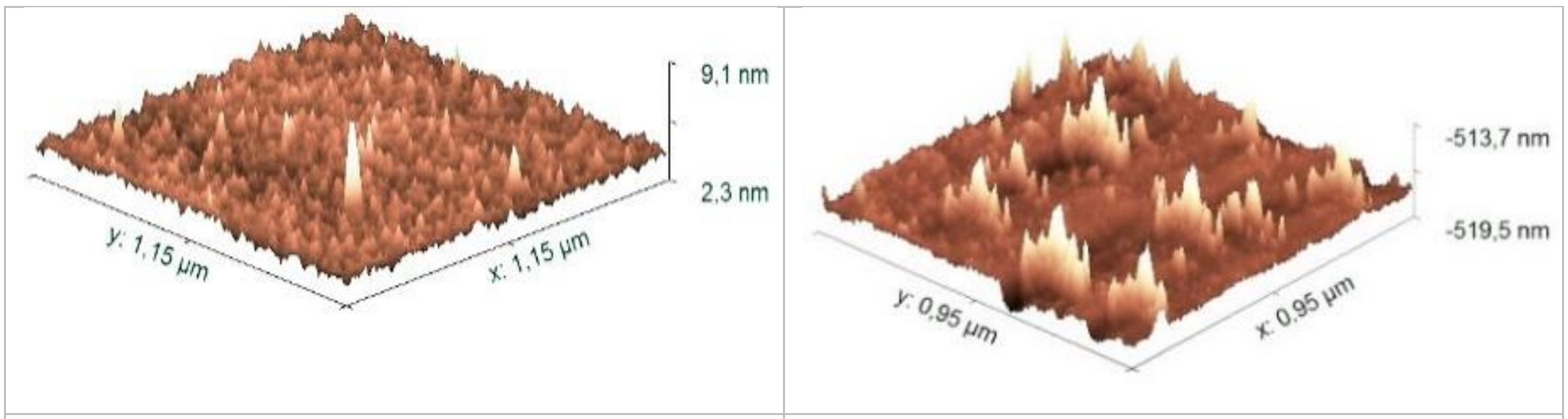

| | |
|---|---|
| (a) C-S6 на Si(100) (500ºC, 10 min) ~ 1.15×1.15, μm | (b) C-S7 на Si(1110-KES) (500ºC, 15 min) ~ 0.95 × 0.95, μm |

Fig. 7(a, b). 3d-images of the layer C-S6 (a) and C-S7 (b) surfaces deposited at $T_g$ = 500°C. The deposition times $t_g$ of: (a) – 10 min; (b) – 15 min.

2d- and 3d-images of the layer C-S5 surface, Fig. 6 (a), show rare high "peaks" (drops of material) with $h_{max}$ ~ 60 nm, which make it difficult to detect small nuclei. The estimate of the thickness of this layer $d_{ell}$ ~ 45 nm. Layer C-S6, Fig. 6 (b): there are rare "peaks" on the surface, the height of which, $h_{max}$ ~ 13 nm, is significantly less than the height of the "peaks" on the C-S5 surface layer. In Fig. 7 (a), fine nuclei covering the entire scanned area are clearly visible. The thickness of the C-S6 layer, $d_{ell}$ ~ 117 nm, significantly exceeds the maximum height of its roughness. Therefore, at $t_g$ = 10 min, a continuous C-layer is formed, on the surface of which n fineuclei continue to form (2D+3D growth mode [17]). On the layer C-S7 surface, Fig. 6(c), there are no peaks, and small nuclei remain. In Fig. 7(b), one can observe coalescence of the nuclei, an increase in their lateral dimensions, and the parallel alignment of larger nuclei. For the layer C-S7, $h_{max}$ of the nuclei is ~ 13 nm, and the thickness $d_{ell}$ ~ 180 nm, i.e., at $t_g$ = 15 min, the layer continues to grow via the 2D+3D mode. The parallel orientation of the nuclei planes (Fig. 7(b)) may be related to the crystallographic orientation of the Si(1110-KES) substrate, but further research is required to clarify the role of substrate orientation in the formation of these layers.

## 4. Conclusion

The optical properties and surface topography of the carbon layers formed by pulsed laser deposition (PLD) on Si substrates were investigated with the growth temperatures ($T_g$) ranging from 200 to 500°C and the deposition times ($t_g$) from 2 to 15 min. The techniques employed included spectroscopic ellipsometry, Raman spectroscopy, atomic force microscopy, and X-ray diffraction reflectometry. Estimates of the carbon layer thicknesses ($d$ ~ 11–180 nm) were obtained, as well as the optical constants spectra, $n(\lambda)$ and $k(\lambda)$, for layers formed at $T_g$ = 500°C with the deposition times of 10 and 15 min ($d$ ~ 117 and 180 nm, respectively). It was determined that at $T_g$ = 500°C, the resulting layers consist of nanocrystalline graphite/4–5-layer graphene with a grain size of approximately ~ 7 nm.

It has been found that at the initial growth stage ($t_g$ = 2 min), the surface topography of the C-layers depends on both the growth temperature ($T_g$) and the crystallographic orientation of the substrates. At $T_g$ = 400°C, nucleation particles with increased lateral dimensions, high density, and a height of ~11 nm forms on the Si(100) substrate; consequently, the temperature range near 400°C is promising for identifying optimal conditions for PLD technology. It has been established that, at $T_g$ = 500°C and deposition times ($t_g$) of 10 and 15 min, a continuous C-layer forms, with small nuclei ($h$ ~ 13 nm) continuing to grow on the surface—a manifestation of a 2D+3D growth mode. Achieving a transition to purely layer-by-layer growth requires longer deposition times.

## Funding

The studies with the use of spectral ellipsometry and atomic force microscopy were carried out as part of a State Task of V.A. Kotelnikov Institute of Radio Engineering and Electronics, Russian Academy of Sciences.